\documentstyle[prl,twocolumn,aps,floats,epsf,epsfig,amsmath,amssymb]{revtex}

\begin{document}

\twocolumn[
\hsize\textwidth\columnwidth\hsize\csname@twocolumnfalse\endcsname

\title{Similarities and Differences in 2D `metallicity' induced by
  temperature and parallel magnetic field: To screen or not to screen}
\author{S. Das Sarma and E. H. Hwang}
\address{Condensed Matter Theory Center, Department of Physics, University of
Maryland, College Park, Maryland 20742-4111}
\date{\today}
\maketitle

\begin{abstract}
We compare the effects of temperature and parallel magnetic field on
the two-dimensional metallic behavior within the unified
model of temperature and field dependent effective disorder arising
from the screened charged impurity scattering.
We find, consistent with experimental observations, that the temperature
and field dependence of resistivity should be qualitatively similar
in n-Si MOSFETs and  different in n-type GaAs 2D systems, with the
p-type 2D GaAs system being somewhat intermediate. Based on our
calculated results we critically comment on the expected similarities
and differences between temperature and field dependent carrier
transport properties in various dilute 2D semiconductor systems. 

\end{abstract}

\vspace{0.5cm}
]

\section{introduction}

It has been experimentally well-established over the last ten years
that low-density and low-disorder two-dimensional (2D) electron systems
often exhibit unusually strong temperature (T) and in-plane (parallel
to the 2D layer) magnetic field ($B_{\|}$) dependence in its low-temperature
($T \sim 1K$) resistivity, $\rho$ \cite{review}. This anomalously
strong temperature and 
field dependence of 2D resistivity in the putative metallic phase,
i.e., for carrier densities $n>n_c$ where $n_c$ is the critical
density for the 2D metal-insulator transition (MIT), i.e. the system
is an effective 2D metal for $n>n_c$ and a strongly localized
insulator for $n<n_c$, has attracted a
great deal of attention because no such behavior is seen in normal 3D
metals where the low-temperature $\rho$ typically saturates (the
so-called Bloch-Gr\"{u}neisen behavior) manifesting little temperature
and/or magnetic field dependence. The purpose of this paper is to
theoretically explore the connection, if any, between $\rho(T)$ and
$\rho(B_{\|})$, i.e., the relationship between temperature and 
magnetic field
dependence of 2D resistivity, and to investigate whether both
originate from a single physical mechanism.

The connection between $\rho(T)$ and $\rho(B_{\|})$ has been emphasized in
the literature, for example, Pudalov {\it et
  al}. \cite{pudalov02} carried out a 
detailed experimental analysis comparing effects
of temperature and parallel field on the strength of the `metallic'
resistivity in Si MOSFETs. 
In particular, Pudalov {\it et al}. found that
``the data for magneto- and temperature dependence of the resistivity of
Si MOS samples in parallel field may be well-described by a simple
mechanism of the magnetic field dependent disorder'' \cite{pudalov02}.
We argue in this paper
through explicit calculations of $\rho(T,B_{\|})$ that this simple
single mechanism giving rise to strong temperature and parallel-field
dependence of 2D resistivity is electronic screening of the quenched
disorder which, for low-density and low-temperature semiconductor
structures of relevance in the 2D MIT problem, arises from the
randomly distributed (unintentional) background charged impurity
centers providing a long-range {\it bare} Coulombic disorder with the
main resistive low-temperature scattering mechanism being carrier
scattering by the effective screened Coulombic impurity disorder.

The physical origin of the
anomalously strong temperature and magnetic field dependence is the
relatively low carrier density involved in the 2D ``metallic'' phase,
which makes it possible for the relevant dimensionless temperature,
$t\equiv T/T_F \propto n^{-1}$, and magnetic field, $b \equiv B_{\|}/B_s
\propto 
n^{-1}$, parameters to be large (of order unity) in the $0.1-1$ K
temperature and $1-10$ T magnetic field range in which the 2D MIT
experiments are typically carried out (where $T_F \equiv E_F/k_B$, the
Fermi temperature, and $B_s = 2E_F/g\mu_B$, the spin polarization
field at which the applied field completely spin-polarizes the 2D
system, 
with $k_B$, $E_F$, $g$, $\mu_B$ being respectively the Boltzmann
constant, the Fermi energy, the effective carrier Land\'{e}
$g$-factor, and Bohr magneton).
Such large values of $t$ and $b$ in 2D semiconductor systems (by
contrast, the typical value of $t$ and $b$ in 3D metals is $10^{-4}$) 
make the screened effective disorder strongly
temperature and magnetic field dependent, leading to unusually strong
temperature and magnetic field dependence of $\rho(T,B_{\|})$.

This `simple' single mechanism of screening
(i.e. strongly temperature and field
dependent effective disorder) also immediately
explains why one needs rather high-quality low-disorder 2D systems
with high mobilities for the manifestation of 2D MIT -- for highly
disordered systems the typical critical density $n_c$ for the 2D MIT
is rather high, making the effective metallic phase exist only at high
carrier densities and therefore the parameter $t$ (and $b$) cannot
become very large (since $t$, $b \propto n^{-1}$) until one reaches
higher temperatures where phonon scattering becomes effective. But,
screening being operational at all carrier densities,
clear signatures of strong temperature and magnetic field dependence of
$\rho(T,B_{\|})$ should exist even at reasonably high densities except that
the total variation $\Delta \rho$ due to increasing $T$ and/or $B_{\|}$
cannot be very large in magnitude at higher densities since the
dimensionless parameters $t$ and $b$ cannot really become large
at high carrier densities. Indeed, such signatures of 2D
``metallicity'' in the temperature dependent resistivity 
are seen in high-quality
Si MOS systems at higher carrier densities. The new feature discovered
by Kravchenko {\it et al.} \cite{Kravchenko94} is the continuous
enhancement of the 
temperature dependence, starting deep in the high-density ``metallic''
phase, as the density is lowered staying within the metallic phase,
arising from the increasing value of $t\equiv T/T_F$ until the
critical density for the insulating phase is reached. Similarly, the
parallel field dependence of 2D resistivity also persists well into
high carrier densities ($n \gg n_c$) deep into the metallic phase
except that the effect becomes quantitatively weaker at higher
densities since the dimensionless $b \equiv B_{\|}/B_s \propto n^{-1}$
cannot become  large at high densities. It is the continuous
density induced evolution of $\rho(T)$ and $\rho(B_{\|})$, as effective $t$
and $b$ acquire large values with decreasing carrier density, which
signals the fact that the physical mechanism responsible for the
anomalously strong temperature/field dependence of $\rho(T,B_{\|})$ at low
densities is already operational at high carrier densities.

The rest of this paper is organized as follows. In sec. II we provide
the theory; in sec. III we give our calculated results for the
magnetoresistance $\rho(B_{\|})$ comparing it with $\rho(T)$ for Si
MOSFETs, 2D n-GaAs and 2D p-GaAs heterostructure systems; in sec. IV
we discuss our results emphasizing the relative importance of
spin-polarization induced magneto-screening suppression and
magneto-orbital coupling effects on the $\rho(B_{\|})$ in different 2D
systems. We conclude in sec. V emphasizing our new results and
qualitative conclusion.

\section{theory}

To understand the $\rho(T,B_{\|})$ behavior quantitatively we start with
the Drude-Boltzmann semiclassical formula for 2D transport limited by
screened charged impurity scattering \cite{dassarma1}:
\begin{equation}
\rho^{-1} \equiv  \sigma = ne^2 \langle \tau \rangle/m,
\end{equation}
where 
\begin{equation}
\langle \tau \rangle =\frac{\int \frac{d^2k}{(2\pi)^2} \epsilon_{\bf k}
  \tau(\epsilon_{\bf k}) \left ( -\frac{\partial f(\epsilon_{\bf k})}
  {\partial \epsilon_{\bf k}} \right )} 
{\int \frac{d^2k}{(2\pi)^2} \epsilon_{\bf k}
   \left ( -\frac{\partial f(\epsilon_{\bf k})}{\partial
  \epsilon_{\bf k}} \right )},
\end{equation}
with the momentum-dependent transport relaxation time
$\tau(\epsilon_{\bf k})$ given by the leading-order approximation:
\begin{eqnarray}
\frac{1}{\tau(\epsilon_{\bf k})} = \frac{2\pi}{\hbar} \int
\frac{d^2{\bf k'}}{(2\pi)^2} \int & &N_i(z)dz \left | u_i({\bf
    k-k'};z) \right |^2 \nonumber \\
& &(1-\cos \theta_{{\bf kk'}}) \delta(\epsilon_{\bf k} -
\epsilon_{{\bf k'}}).
\end{eqnarray}
In Eqs. (1)-(3) $\sigma$ and $\tau$ are respectively the 2D
conductivity and relaxation time with $\epsilon_{\bf k}$,
$f(\epsilon_{\bf k})$ as
the electron energy [$\epsilon_{\bf k}=\hbar^2k^2/2m$] and the Fermi
distribution function respectively. Eqs. (1) and (2), defining the
conductivity and the thermally averaged relaxation time respectively,
are standard results of the Boltzmann transport theory whereas Eq. (3)
defining the momentum-dependent relaxation time explicitly depends on
the three-dimensional distribution of charged impurities, $N_i(z)$.
The most important quantity for our consideration is
the effective disorder scattering potential $u_i({\bf q};z)$ in
Eq. (3), which, for charged impurity scattering, should be the
screened charged impurity scattering:
\begin{equation}
u_{i}({\bf q};z) \equiv v_i({\bf q};z)/\varepsilon(q),
\end{equation}
where $v_i$ is the bare (Coulomb) electron-impurity scattering
potential, and $\varepsilon(q)$ is the static RPA dielectric
function. The strong variation in $\rho(T;B_{\|})$ with $T$ and
$B_{\|}$ arises in this 
simple physical picture from the strong variation in the effective
disorder potential $u_i$ due to the variation in the dielectric
(i.e. the screening) function $\varepsilon(q)$ imposed by large
variations in the dimensionless temperature ($t\equiv T/T_F$) and
magnetic field ($b \equiv B_{\|}/B_s$). Note that at $T=0$, we have
$\varepsilon(q) = 1 + 
q_{TF}/q$ in 2D for $q \le 2k_F$, and $2k_F$ scattering, $|{\bf k-k'}|
= 2k_F$, is the most important resistive scattering process so that
$q_{TF}/2k_F \propto n^{-1/2}$ is the nominal control
parameter for the strength of metallicity.
The Thomas-Fermi screening wave vector $q_{TF}$ and the Fermi wave
vector in 2D systems are given respectively by the formula
$q_{TF}=g_dme^2/(\kappa \hbar^2)$ and $k_F = (4 \pi n/g_d)^{1/2}$, where
$n$ is the 2D carrier density, $\kappa$ is the background lattice
dielectric constant, and $g_d$ is the degeneracy factor including the
spin degeneracy.

Both of these  situations, screening-induced variations in
$\rho(T)$ \cite{dassarma1} and in $\rho(B_{\|})$ \cite{gold_b}, have
already been separately theoretically 
discussed in the 2D MIT literature, with considerable success in
qualitatively explaining the experimental data. The purpose of the
current paper is to critically analyze the connection and 
relationship between the overall strengths (``metallicity'') of
temperature and field dependence of resistivity. 
In particular, the important
effect of the in-plane field on screening is through
spin-polarization of the carriers which could continuously change the
spin-degeneracy factor from 2 (at $B_{\|}=0$) to 1 (at $B_{\|}=B_s$,
when the 
carriers are completely spin-polarized), so that the Thomas-Fermi
screening wave vector $q_{TF}$, which is proportional to the density
of states and hence to the spin degeneracy factor $g_s$, decreases by
a factor of 2 as $B_{\|}$ goes from $B_{\|}=0$ to $B_{\|}\ge
B_s$. {\it In the strong 
  screening situation} (i.e., for $q_{TF} \gg 2k_F$, where $k_F
\propto \sqrt{n}$ is the Fermi wave vector), which typically applies
to Si MOSFETs (but {\it not} to 2D n-GaAs systems), this would then
imply that $\rho(B_{\|})/\rho(B_{\|}=0) \equiv r_b(B_{\|})$ could have
an absolute maximum value  
of 4 in the ideal situation with $r_b(B_{\|})$ being a constant for
$B_{\|}>B_s$, and  $r_b(B_{\|})$ rising continuously in the $0\le
B_{\|}\le B_s$ 
regime. Si MOSFETs seem to obey this behavior quite well since $q_{TF}
\gg 2k_F$ strong screening condition is well satisfied in Si MOSFETs
at all experimental ``metallic'' densities ($n\ge 10^{11}cm^{-2}$).

In discussing $\rho(T)$,
however, it is not possible to obtain such a simple maximum ideal
value for $\rho(T)/\rho(T=0) \equiv r_t(T)$ because thermal effects on
screening do {\it not} saturate at some optimum temperature (unlike
the magnetic field induced spin polarization dependent screening effect
which does saturate at $B_{\|}=B_s$). In
principle, screening vanishes in the extreme high temperature limit,
$T \gg T_F$, leading to an absolute upper limit on the temperature
induced resistivity enhancement of $\rho(T\gg T_F)/\rho(T=0) \equiv
r_t(T \gg T_F) \approx (q_{TF}/2k_F)^2$ through the screening effect
(in the strong screening $q_{TF} \gg 2k_F$ regime). We emphasize that,
unlike $r_s(B_{\|})$ which actually does have a reasonable ideal maximum of
4 for magnetic field induced resistivity enhancement, the {\it
  theoretical} ideal maximum $r_t(T\gg T_F) = (q_{TF}/2k_F)^2$ is {\it
  not} a reasonable value for the practical observable temperature
induced maximum enhancement of resistivity. This is because at 
high temperatures ($T \sim T_F$) various other effects (e.g. 
smearing of the Fermi surface, thermal averaging, etc.) become
operational and 
$\rho(T)$ {\it actually} decreases \cite{dassarma1}
with increasing temperature in the
$T \ge T_F$ regime. 
(Also, at higher temperatures phonon scattering invariably becomes
significant, leading eventually to a monotonically increasing $\rho(T)$
with temperature.)
Nevertheless, $r_t(T\gg T_F) = (q_{TF}/2k_F)^2$ is
{\it not} completely wrong as a crude qualitative measure
of the theoretical upper limit on the temperature induced maximal
enhancement of $\rho(T)$. It is easy to derive a simple formula
connecting $\tilde{r}_b \equiv r_b(B_{\|}=B_s)$ and $\tilde{r}_t \equiv
r_t(T/T_F 
\rightarrow \infty)$ {\it using simply the screening considerations}
given above (which will obviously not be correct, but may be used for
qualitative considerations): $\tilde{r}_b \equiv 4 \tilde{r}_t
[\tilde{r}_t + 2 
\sqrt{\tilde{r}_t} + 1]^{-1}$. We caution that this formula typically 
strongly overestimates $r_t$.
The basic picture that emerges from the above qualitative
considerations is that $\rho(T)$ and $\rho(B_{\|})$ should, in general,
have ``similar'' anomalously strong metallic behavior -- in
particular, in the strong screening situation ($q_{TF} \gg 2k_F$)
both should show anomalously large increase in the resistivity as $t$
and $b$ increase from zero to order unity (which is made possible by
low Fermi energy at low carrier densities so that $t\equiv T/T_F$ and
$b \equiv B_{\|}/B_s$ can be large). Note that at this stage
it is not possible to make a more direct quantitative connection
(except for the simple formula which overestimates the temperature
dependence) between $r_t$ and $r_b$ except to say that they should
both be large or both be small.
Note also that we have only discussed so far the spin-polarization
induced magneto-screening effect on $\rho(B_{\|})$ leaving out other
possible corrections (e.g. magneto-orbital coupling) which could be
quantitatively important.

\section{results}

We now apply these
concrete considerations to n-Si MOSFET and n-GaAs 2D systems, which
are representative examples of strong-screening (n-Si MOSFET)
and weak-screening (n-GaAs) 2D systems, respectively.  We also show
calculated results for holes in the 2D p-GaAs system which is somewhat
intermediate between Si MOSFETs and n-GaAs systems in terms of
screening properties.

\begin{figure}
\epsfysize=1.9in
\centerline{\epsffile{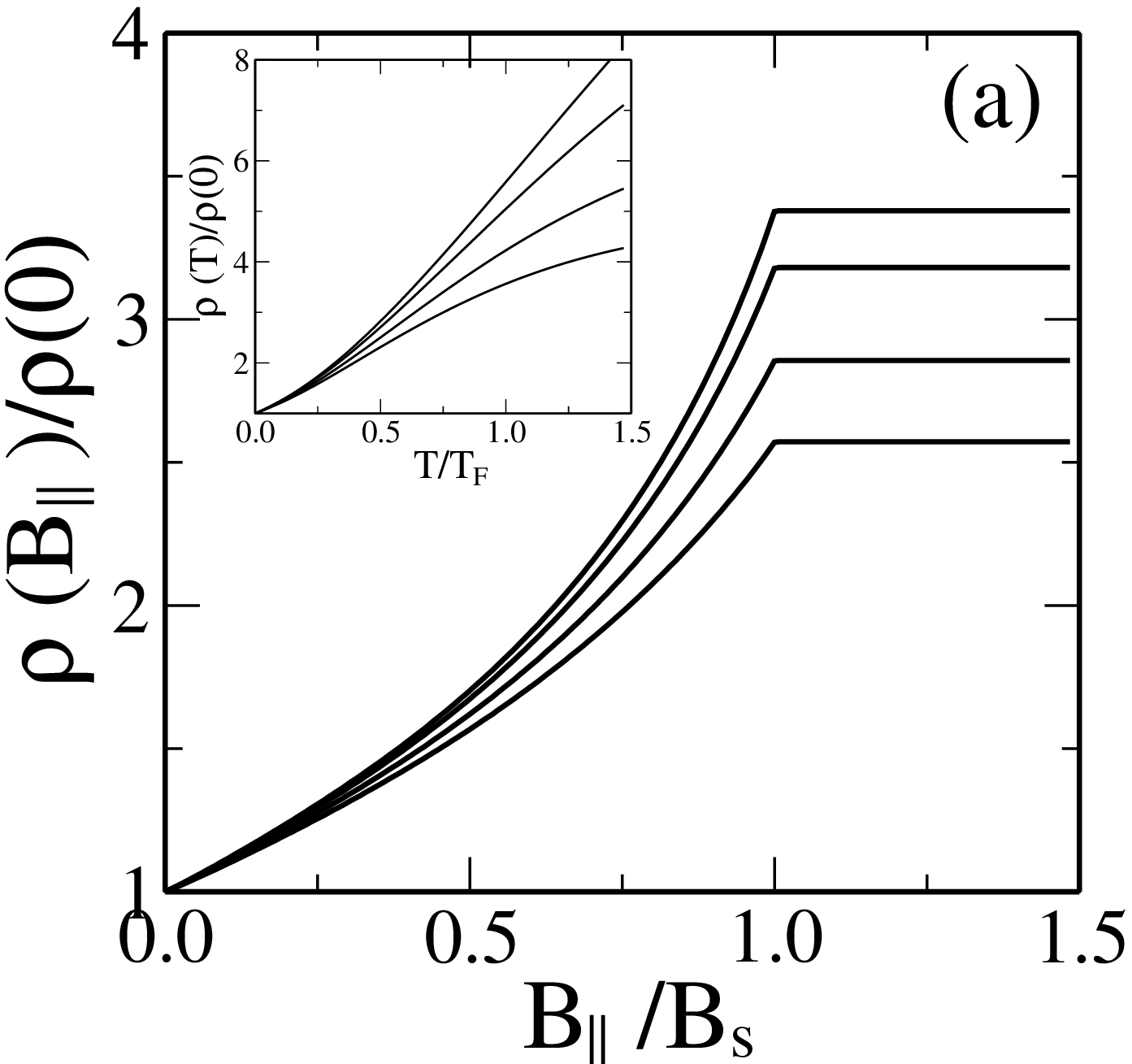}}
\epsfysize=1.9in
\centerline{\epsffile{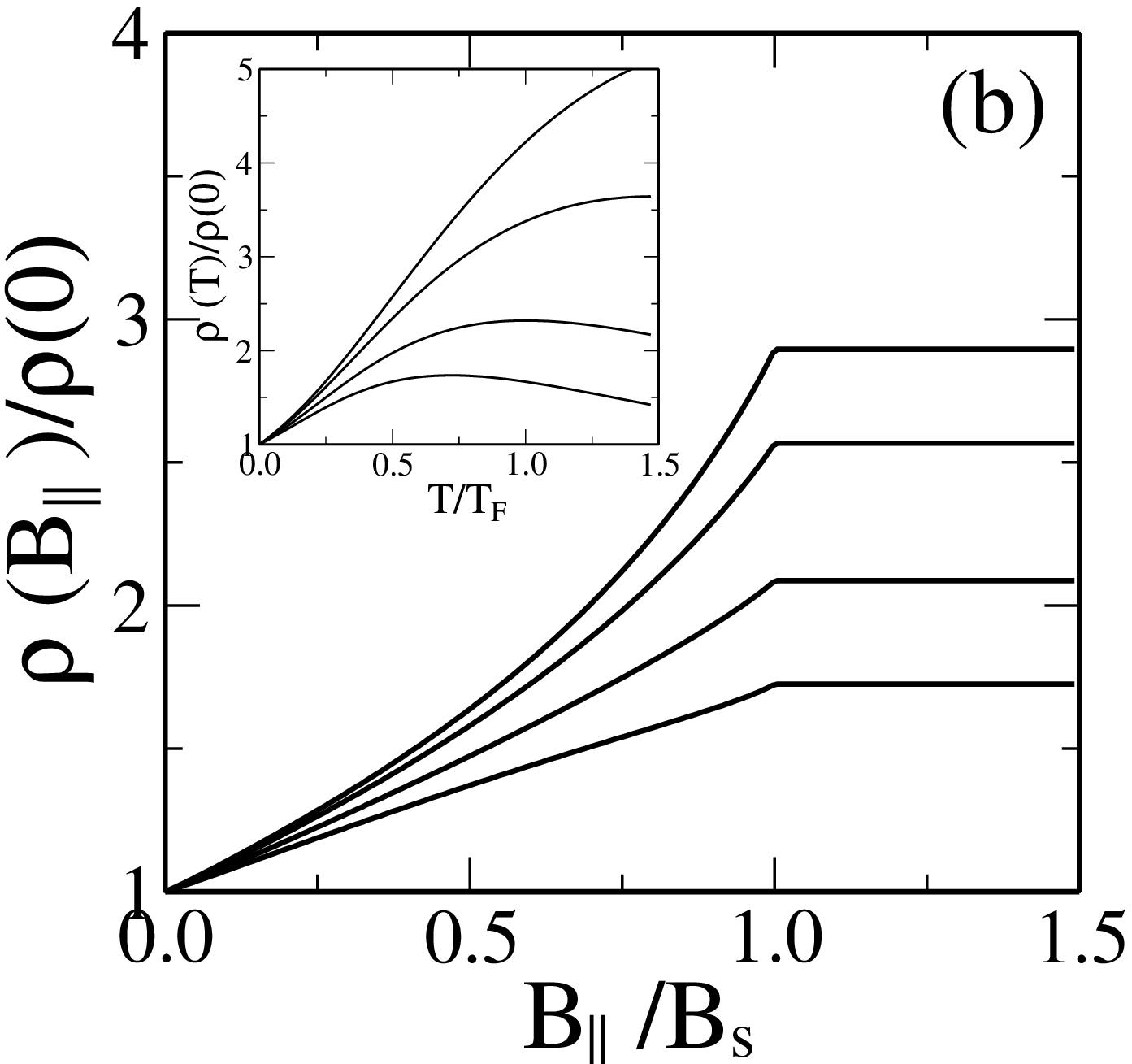}}
\caption{
The calculated $\rho(B_{\|})$ in Si MOSFETs (a) for ideal 2D system
and (b) for quasi-2D system at $T=0$ and for various densities
$n=1.0$, 2.0, 5.0, 10.0$\times 10^{11} cm^{-2}$ (from top to bottom). 
The insets show 
$\rho(T)$ at $B_{\|} =0$ and for the same systems as the main figures.
}
\label{fig_Si}
\end{figure}

In Fig. \ref{fig_Si} we show our full numerical Drude-Boltzmann
calculations for 
$\rho(T)$ and $\rho(B_{\|})$ in Si MOSFETs assuming ideal (zero-width) 2D
and (finite-width) quasi-2D systems with the charged impurity
scatterers located randomly at the Si-SiO$_2$ interface. It is clear
that for Si MOSFETs $\rho(T)$ and $\rho(B_{\|})$ manifest qualitatively
similar ``anomalous'' metallic behavior over a large range of
density, temperature, and magnetic field values, both for the strictly
2D and for the realistic quasi-2D models. 
This qualitative similarity,
already noted in the corresponding experimental situation by Pudalov
{\it et al.} \cite{pudalov02}, arises entirely from the strong-screening
nature ($q_{TF} 
\gg 2k_F$) of Si MOSFETs due to its rather large effective mass and
valley degeneracy of 2 which makes $q_{TF}/2k_F|_{\rm Si} \approx 16
/\sqrt{\tilde{n}}$ where $\tilde{n}$ is the carrier density measured
in units 
of $10^{11}cm^{-2}$. In the usual density range of our interest,
therefore, the strong screening condition $q_{TF} \gg 2k_F$ is always
satisfied for Si MOSFETs leading to qualitatively similar behavior in
$\rho(T)$ and $\rho(B_{\|})$, as shown in Fig. \ref{fig_Si} and as
observed experimentally.

\begin{figure}
\epsfysize=1.9in
\centerline{\epsffile{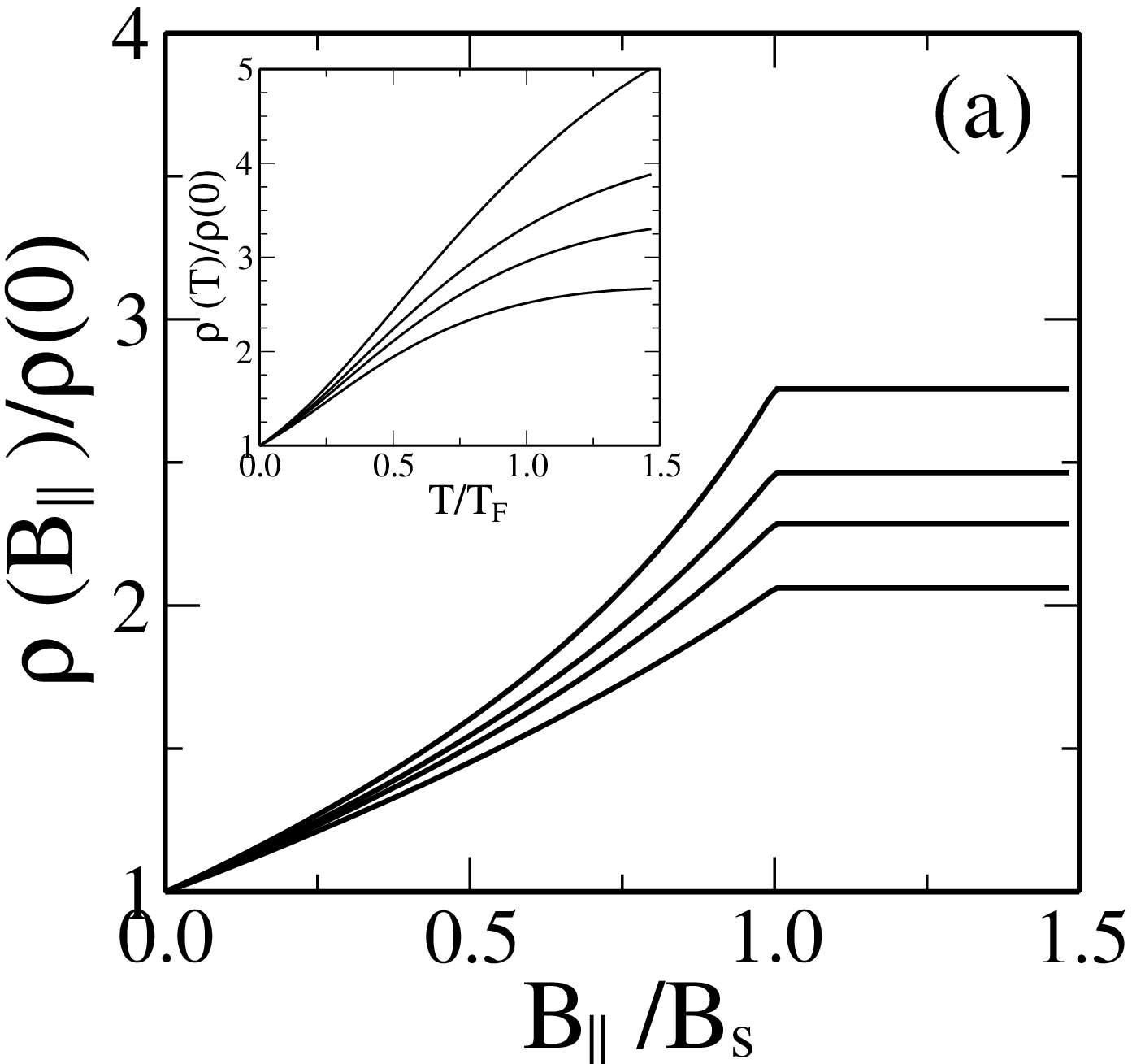}}
\epsfysize=1.9in
\centerline{\epsffile{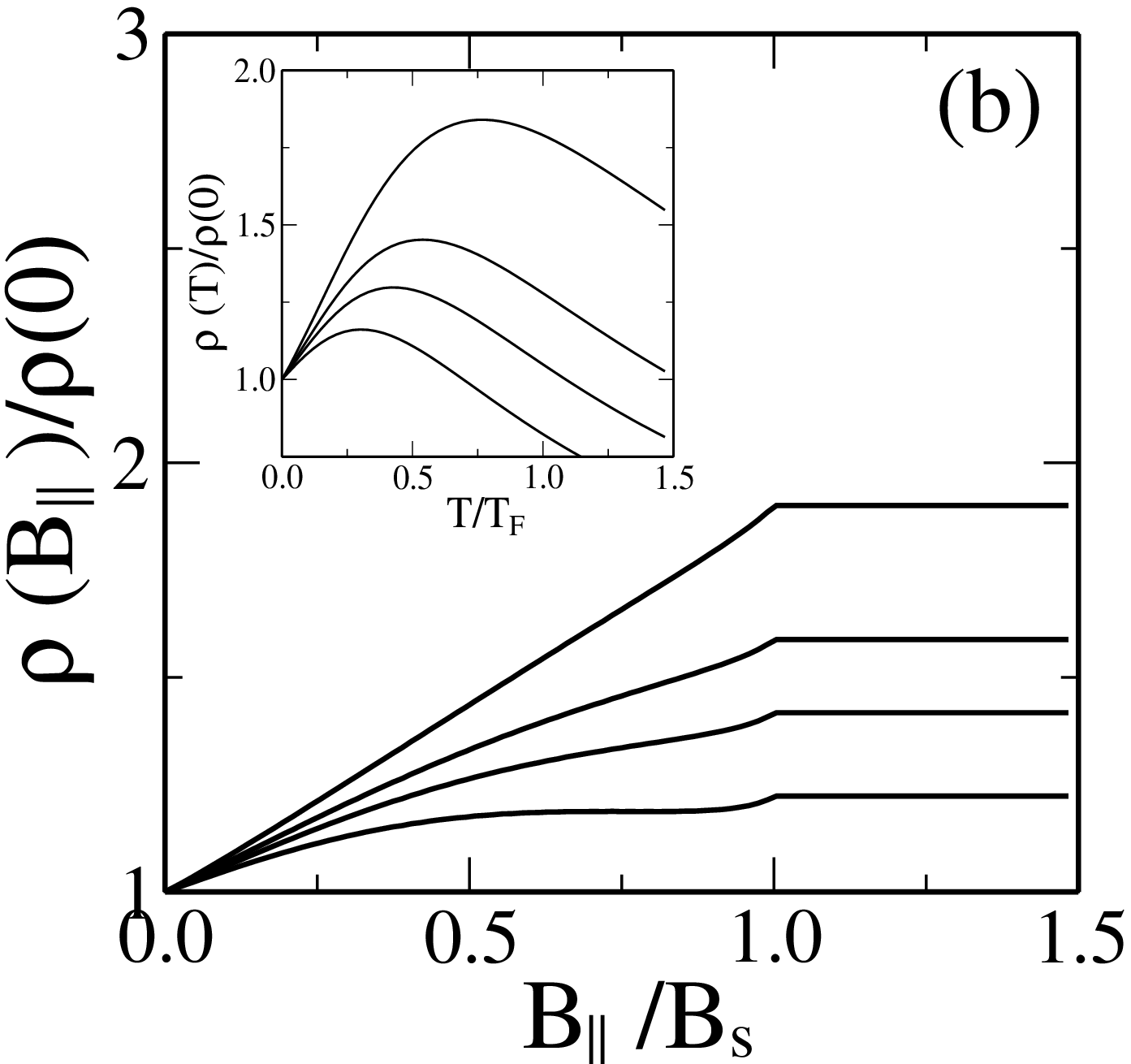}}
\caption{
The calculated $\rho(B_{\|})$ in n-GaAs (a) for ideal 2D system
and (b) for quasi-2D system at $T=0$ and for various densities
$n=0.4$, 0.8, 1.2, 2.0$\times 10^{10} cm^{-2}$ (from top to bottom). 
The insets show 
$\rho(T)$ at $B_{\|} =0$ and for the same systems as the main figures.
}
\label{fig_nGaAs}
\end{figure}

Now we consider the opposite extreme of electrons in 2D n-GaAs
heterostructure system where screening is generally weak and the
$q_{TF} \gg 2k_F$ strong-screening condition can only be satisfied at
very low carrier densities: $q_{TF}/2k_F|_{\rm n-GaAs} \approx
1.3/\sqrt{\tilde{n}}$. We therefore expect rather weak temperature
and field dependence of resistivity in 2D n-GaAs system, which is even
further exacerbated by the relatively large values of the Fermi energy
(due to the small effective mass) which make the relative
values of the dimensionless temperature ($t\equiv T/T_F$) and magnetic
field ($b\equiv B_{\|}/B_s$) parameters rather low in the experiments. In
Fig. \ref{fig_nGaAs} we show our 2D n-GaAs calculation for $\rho(T)$ and
$\rho(B_{\|})$, again in the ideal 2D and realistic quasi-2D
approximations, taking into account screened charged impurity
scattering at the interface. As expected (due to the relatively low
value of $q_{TF}/2k_F$ in n-GaAs), both temperature and magnetic field
dependence are weak in Fig. \ref{fig_nGaAs}, again manifesting the
underlying 
connection between the field and temperature dependence of the
effective disorder arising from the temperature and magnetic field
dependent screening of charged impurity screening.

\begin{figure}
\epsfysize=2.0in
\centerline{\epsffile{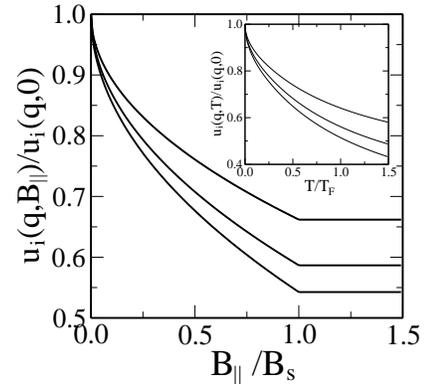}}
\caption{
The calculated screened effective 
disorder scattering potential $u_i(q,B_{\|})/u_i(q,0)$ at $q=2k_F$ and
T=0 as a function 
of the magnetic field for various screening
parameter $q_{TF}/2k_F = 5$, 10, 20 (from top to bottom). The inset
shows $u_i(q,T)/u_i(q,0)$ at $q=2k_F$ and B=0 as a function of
temperature. 
}
\label{fig_eff}
\end{figure}

Comparing Figs. \ref{fig_Si} and \ref{fig_nGaAs} we tentatively
conclude that there is a 
compelling theoretical qualitative similarity between $\rho(T)$ and
$\rho(B_{\|})$ for charged impurity scattering limited 2D transport arising
essentially entirely from the ``similar'' temperature and field
induced weakening of screening: strong (weak) variation in
$\rho(T)$ implies corresponding strong (weak) variation in $\rho(B_{\|})$
and vice versa as is manifest in Fig. \ref{fig_Si}(\ref{fig_nGaAs})
for n-Si MOSFET 
(n-GaAs). To make this point more explicit we show in
Fig. \ref{fig_eff} our 
calculated screened effective impurity 
disorder $|u_i(q=2k_F)|$ as a function
of the dimensionless magnetic field and temperature 
parameters for a number of different values of the screening
parameter $q_{TF}/2k_F$. It is obvious that for large values of
$q_{TF}/2k_F$, the effective disorder varies strongly (and similarly)
with $t$ and $b$, leading to qualitatively similar strong variation in
$\rho(T)$ and $\rho(B_{\|})$.

The above-discussed qualitative theoretical similarity in $\rho(T)$
and $\rho(B_{\|})$ is entirely consistent with the available experimental
data in Si MOSFETs, where both $\rho(B_{\|})$ and $\rho(T)$ show
anomalously strong variations at low carrier densities. But, in 2D
n-GaAs system the existing experimental data on $\rho(T)$ and
$\rho(B_{\|})$ are in {\it striking disagreement} with the screening theory
predictions shown in our Fig. \ref{fig_nGaAs}. In particular, the
experimental 
$\rho(B_{\|})$ shows a strong variation with the magnetic field whereas the
experimental $\rho(T)$ shows a rather weak variation \cite{zhu,lilly},
which obviously 
is inconsistent with the simple screening picture. While a part of
this discrepancy can be explained by the experimental value of the
dimensionless temperature parameter (i.e., $t\equiv T/T_F$) being
smaller than the corresponding magnetic field parameter (i.e.,
$b\equiv B_{\|}/B_s$) because phonon scattering becomes important at higher
temperatures in GaAs, the overall qualitative disagreement between
experiment and theory in 2D n-GaAs system is inexplicable
within the screening theory. Some other ingredient of physics is
missing in the theory as far as the $\rho(B_{\|})$ behavior is concerned in
the 2D GaAs system -- the $\rho(T)$ behavior is qualitatively
well-explained \cite{lilly} by the screening theory results shown in
Fig. \ref{fig_nGaAs}. The 
missing piece of physics is the magneto-orbital effect
\cite{DH6} which affects the 2D GaAs system much more 
strongly than the 
2D Si system since the 2D confinement is much weaker in GaAs by virtue
of its small effective mass and low density. When the quasi-2D layer
width is larger than the magnetic length associated with the parallel
field, the 2D orbital dynamics is strongly affected by the magnetic
field with a strong increase in the field dependent 2D effective mass
as well as by intersubband scattering \cite{DH6}. 
We refer to our earlier work \cite{DH6} for details on the
magneto-orbital effects.

\begin{figure}
\epsfysize=2.0in
\centerline{\epsffile{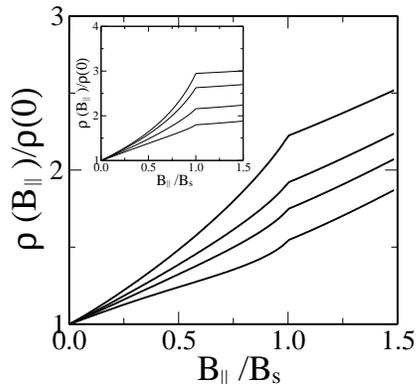}}
\caption{
The main figure (inset) shows the calculated $\rho(B_{\|})$ in the
realistic quasi-2D n-GaAs (Si MOS) 
system including both magneto-spin polarization and magneto-orbital
effects for various densities $n=0.4$, 0.8, 1.2, 2.0$\times 10^{10}
cm^{-2}$ ($n=1.0$, 2.0, 5.0, 10.0$\times 10^{11} cm^{-2}$), from top
to bottom. 
}
\label{fig_orbit}
\end{figure}

In Fig. \ref{fig_orbit} we show our
calculated $\rho(B_{\|})$ in the realistic quasi-2D n-GaAs (and Si MOS)
system including both magneto-spin polarization and magneto-orbital
effects \cite{DH6}. It is clear that both magneto-spin
polarization and 
magneto-orbital effects are essential in understanding the GaAs data
whereas in Si MOSFETs the magneto-orbital effects are small due to the
rather tight confinement of the quasi-2D electron wavefunction. We
note that $\rho(B_{\|}>B_s)$ continues increasing in n-GaAs because of
the magneto-orbital effect whereas the spin polarization effect saturates
at $B_{\|}=B_s$. We emphasize that the results of Fig. \ref{fig_orbit}
are in qualitative agreement with the experimental $\rho(B_{\|})$ in
ref. \onlinecite{zhu} 
whereas the theoretical results with just the spin polarization
effects disagree with experiment. One therefore needs to
include both magneto-screening and magneto-orbital effects for
understanding 2D n-GaAs heterostructure
systems whereas experimental data for Si
MOSFETs (because of their tight 2D confinement) can be reasonably
well-explained without any the magneto-orbital effects.

\begin{figure}
\epsfysize=2.in
\centerline{\epsffile{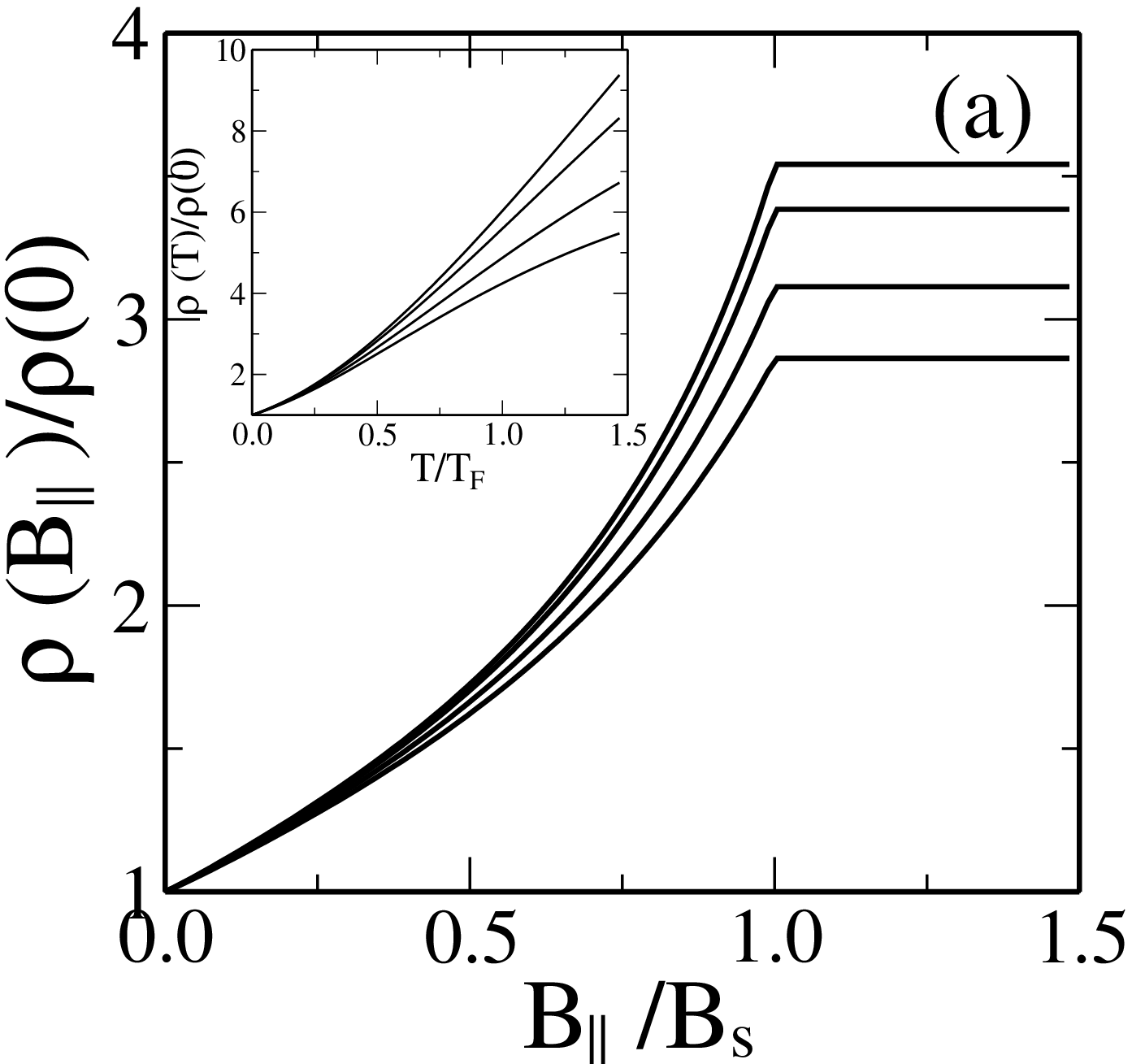}}
\epsfysize=2.in
\centerline{\epsffile{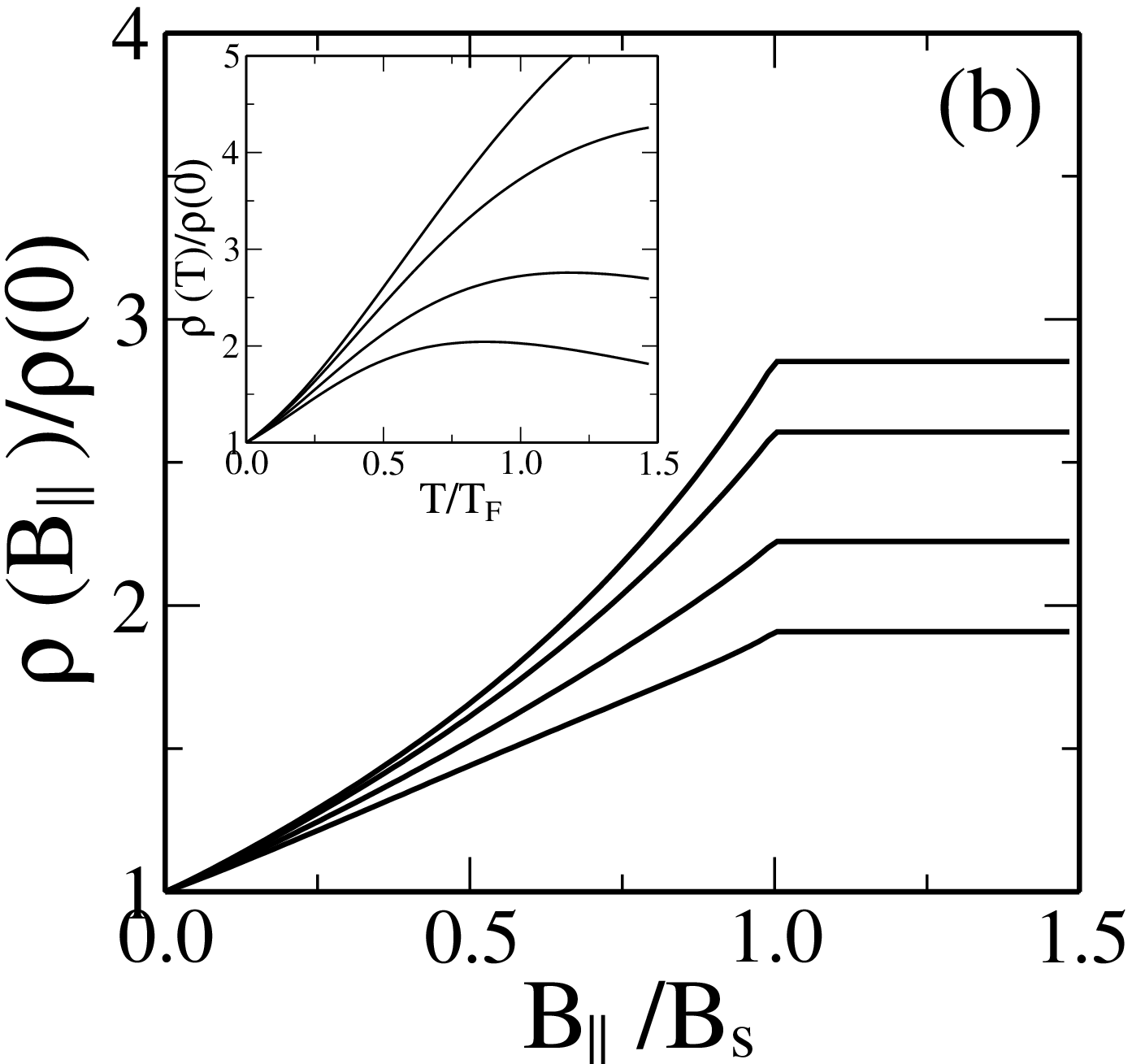}}
\epsfysize=2.in
\centerline{\epsffile{fig5c.eps}}
\caption{
The calculated $\rho(B_{\|})$ in p-GaAs (a) for ideal 2D system
and (b) for quasi-2D system at $T=0$ and for various densities
$n=1.0$, 2.0, 5.0, 10.0$\times 10^{10} cm^{-2}$ (from top to bottom). 
The insets show 
$\rho(T)$ at $B_{\|} =0$ and for the same systems as the main figures.
In (c) the calculated $\rho(B_{\|})$ in the quasi-2D system including
both magneto-spin and magneto-orbital effects is shown for the same
densities given in (b).
}
\label{fig_pGaAs}
\end{figure}

In this context, it is also 
interesting to consider 2D holes in the 
p-GaAs heterostructures because it is a strongly
screening ($q_{TF} \gg 2k_F$) system, which also could have
substantial magneto-orbital coupling at low carrier densities due to
its weak confinement potential leading to
the measured magnetic field dependence of the resistivity being quite
different from that of Si MOSFET systems. 
In Fig. \ref{fig_pGaAs} we show our calculated $\rho(T)$ and
$\rho(B_{\|})$ of the p-GaAs system in the ideal 2D and realistic quasi-2D
systems taking into account screened charged impurity
scattering at the interface. As in the n-Si MOSFET systems, the high
value of $q_{TF}/2k_F$ in p-GaAs
in the density range of our interest ($n \ge
10^{10} cm^{-2}$), $q_{TF}/2k_F|_{\rm p-GaAs} \approx 8 
/\sqrt{\tilde{n}}$ with $m^*_h = 0.4 m_e$,  gives rise to
both strong temperature and magnetic field dependence of the resistivity.
However, the experimental $\rho(B_{\|})$ of p-GaAs shows no saturation
\cite{yoon}
behavior at $B_{\|} \ge B_s$. Again the increase of
$\rho(B_{\|})$ above $B_s$ in p-GaAs heterostructure arises from
the magneto-orbital effect due to the low-density quasi-2D width being
larger than the parallel field induced magnetic length.
In Fig. \ref{fig_pGaAs}(c),
we show $\rho(B_{\|})$ including both magneto-spin and
magneto-orbital effects.

\section{discussion}

The main purpose of this article is to point out that the observed
``similarity'' between the temperature dependence, $\rho(T)$, and the
parallel field dependence, $\rho(B_{\|})$, of the 2D resistivity in Si
MOSFETs arises from the importance of screening of the long-range
Coulombic disorder (associated with random charged impurity centers)
at low carrier densities. In particular, increasing $T/T_F$ reduces
screening effects making the effective disorder stronger, leading to a
large increase in $\rho(T)$. Similarly, increasing the parallel field
reduces the screening since the 2D carriers become spin-polarized by
the applied field, leading to a decreasing density of states and hence
decreasing screening effects. Thus, to the extent screening is the
dominant mechanism (i.e. as long as $q_{TF}/2k_F \gg1$)
underlying 2D metallic behavior, one expects a
qualitative similarity between $\rho(T)$ and $\rho(B_{\|})$ as observed
empirically in ref. \cite{pudalov02} for Si MOSFETs.

Another important point made in this paper is that, although screening
is the dominant physical mechanism underlying 2D metallicity, there
are several other factors which could affect $\rho(T)$ and
$\rho(B_{\|})$, 
leading to considerable difference between the two. In particular,
$\rho(T)$ is affected by phonon scattering and thermal averaging at
``higher'' values of temperature, and $\rho(B_{\|})$ is affected by the
magneto-orbital effect for large quasi-2D widths of the system. While
these non-screening effects are quantitatively unimportant in Si
MOSFETs in the usual $n \sim 10^{11}cm^{-2}$ experimental density
range, the magneto-orbital (as well as phonon scattering and thermal
averaging) effects turn out to be important for n-GaAs 2D
heterostructures where the 2D confinement potential becomes very weak
at low carrier densities leading to strong magneto-orbital
effects. Therefore, in 2D n-GaAs heterostructures (and to a lesser
extent in 2D p-GaAs heterostructures), there are significant
differences between $\rho(T)$ and $\rho(B_{\|})$, compared with the Si
MOSFET situation, arising from magneto-orbital coupling.

We note that the magneto-orbital coupling effect, where the applied
parallel field couples directly to the carrier orbital dynamics in
addition to causing spin-polarization, arises only when the quasi-2D
transverse width $\langle z \rangle$ of the system is larger than the
magnetic length $l_B=\sqrt{\hbar c/eB_{\|}}$ associated with the applied
parallel field. The magneto-orbital effect may be suppressed in n- and
p-GaAs systems by using narrow quantum well systems (with $\langle z
\rangle < l_B$) rather than heterostructure systems. In
heterostructure, the 2D confinement potential is provided by the
self-consistent potential which becomes very shallow at low carrier
densities --- in particular, $\langle z \rangle \propto n^{-1/3}$
where $n$ is the carrier density. In a quantum well, on the other
hand, the width ($a$) of the quantum well completely determines the
strength of the magneto-orbital coupling, and as long as $a <l_B$,
magneto-orbital coupling effects can be ignored. This is indeed the
experimental finding \cite{gao} in p-GaAs quantum wells. We also note
that the dimensionless parameters $B_{\|}/B_s \sim B_{\|}/n$ and
$\langle z \rangle /l_B \sim B_{\|}^{1/2}/n^{1/3}$ respectively
control the strength of magneto-screening and magneto-orbital effects
in 2D systems. In Si MOSFETs $\langle z \rangle \le 50 \AA$ in the
experimental density range, and the magneto-orbital coupling is
unimportant. Finally, we add that the thermal effects on $\rho(T)$ are
controlled by the parameters $T/T_F$ and $q_{TF}/2k_F$ as long as
phonon scattering effects are unimportant --- phonons become important
above $5-10$ K for Si MOSFETs and above $1-2$K for n- and p-GaAs 2D
systems.

\section{conclusion}

Before concluding it may be worthwhile to emphasize what is new in
this paper compared with existing theoretical publications on the
temperature and parallel magnetic field dependence of low-density
carrier transport in 2D semiconductor structures. We have earlier
\cite{dassarma1} extensively reported on theoretical results
describing the temperature dependence of 2D resistivity in low-density
semiconductor structures, and therefore the temperature dependent
$\rho(T)$ results shown as insets of our figures are 
only provided for the
purpose of comparison with the magnetic field dependent resistivity
$\rho(B_{\|})$ shown in the main panels of our figures. Our magnetic field
dependent 2D resistivity, $\rho(B_{\|})$, in different systems is new, and
shows that, as long as the magneto-orbital coupling of the parallel
field to the transverse orbital dynamics of the 2D system can be
neglected (i.e. as long as the quasi-2D transverse width of the system
is small compared with the magnetic length $l_B$), the
suppression of the screening of the impurity potential by the magnetic
field (due to spin polarization) and by the temperature (due to
thermal smearing) is very similar. The two systems, the n- and p-type
GaAs heterostructures, where the magneto-orbital coupling is strong,
the temperature-dependent resistivity $\rho(T)$ and the
field-dependent resistivity $\rho(B_{\|})$ behave differently. It may also
be useful to ask about the importance of interaction effects
\cite{zala} beyond screening in determining the temperature and field
dependent 2D resistivity. While interaction effects are undoubtedly
important, their quantitative significance is unknown at this stage
since the interaction theory appropriate for the realistic long-range
impurity potential does not yet exist. The qualitative agreement
between our screening theory and experimental results indicate that
interaction corrections are of quantitative, but {\it not} of
qualitative importance. One important quantitative effect of
interaction is the actual value of the spin-polarization field,
$B_s=2F_F/g\mu_B \propto n/(gm)$, which is inversely proportional to
the product $gm$. It is known \cite{review,zhu} that $gm$ is strongly
renormalized by interaction effects \cite{zhang}, leading
to a large suppression of $B_s$ compared with the non-interacting value.

We conclude by emphasizing our qualitative findings: (1) in Si MOSFETs,
the parallel field dependence of 2D resistivity (at a constant low
temperature $T/T_F \ll 1$) $\rho(B_{\|})$ and the temperature
dependence of the resistivity (at zero parallel field)
$\rho(T)$ should manifest qualitatively ``similar''
anomalously strong metallic behavior since both
are dominated by screening effects; (2) in n-GaAs $\rho(T)$ should
show weak metallicity by virtue of weak screening, but $\rho(B_{\|})$
in low-density heterostrucutres
should show strong $B_{\|}$-dependence since it is
affected both by spin-polarization and magneto-orbital effects; 
(3) in p-GaAs due to strong screening the resistivity 
shows both strong temperature and magnetic field dependence, and the
magneto-orbital effect is operational in weak confinement 
heterostructure systems;  
(4) the ``strong'' metallicity in 2D semiconductor structures arises
from the possibility of having strong screening (i.e. large
$q_{TF}/2k_F$) 
as well as large values of dimensionless temperature ($T/T_F$) and
magnetic field ($B_{\|}/B_s$), where $B_s$ is the magnetic field for
full spin-polarization of the 2D system; (5) experiments should be
carried out in both low-density 2D heterostructures and narrow quantum
wells (for n- and p-GaAs systems) in order to separate out
magneto-screening and magneto-orbital corrections to $\rho(B_{\|})$.

This work is supported by ONR, NSF, ARO, ARDA, DARPA,
and LPS.

\end{document}